          [2001/04/25 1.1 (PWD)]
\documentclass[a4paper,twocolumn]{esapub} % European paper
\usepackage{natbib,graphicx}

\newcommand{\Al}{$^{26}$Al\ }
\newcommand{\about}{$\simeq$}
\newcommand{\flux}{ph~cm$^{-2}$s$^{-1}$}
\newcommand{\msol}{M$_{\odot}$}
\title{\Al Studies with INTEGRAL's Spectrometer SPI}
\author{R. Diehl, K. Kretschmer, G. Lichti, V. Sch\"onfelder, A.W. Strong, 
A. von Kienlin}
\affil{Max-Planck-Institut f\"ur extraterrestrische Physik, D-85741 Garching,
Germany}
\author{J. Kn\"odlseder, P. Jean, V. Lonjou, G. Weidenspointner, 
J.-P. Roques, G. Vedrenne}
\affil{Centre d'Etude Spatiale des Rayonnements,UPS-CNRS,  31028 Toulouse, 
France}
\author{S. Schanne}
\affil{DSM/DAPNIA/Service d'Astrophysique, CEA Saclay, 91191 Gif-Sur-Yvette,
 France}
\author{N. Mowlavi}
\affil{INTEGRAL Science Data Center, Chemin d'Ecogia, CH-1290 Versoix,
 Switzerland}
\author{C. Winkler}
\affil{ESA/ESTEC, Science Operations and Data Systems Division (SCI-SD)  
2201 AZ Noordwijk, The Netherlands 		  }
\author{C. Wunderer}
\affil{Space Sciences Lab, UC Berkeley, Berkeley, CA 94720, USA}
\begin{document}

\maketitle


\begin{abstract}
\Al radioactivity traces recent nucleosynthesis throughout the Galaxy,
and is known to be produced in massive stars and novae. The 
map from its decay gamma-ray line suggests massive stars to dominate,
but high-resolution line spectroscopy is expected to supplement imaging
of \Al source regions and thus to help decide about
the \Al injection process and interstellar environment, hence about
the specific massive-star subgroup and phase which produces interstellar $^{26}$Al. 
The INTEGRAL Spectrometer SPI has observed Galactic \Al radioactivity
in its 1809 keV gamma-ray line during its first inner-Galaxy survey.
Instrumental background lines make analysis difficult; yet, a clear
signal from the inner Galaxy agrees with expectations. In particular,
SPI has constrained the line width to exclude previously-reported
line broadenings corresponding to velocities $>$500~km~s$^{-1}$.  
The signal-to-background ratio of \about~percent implies that detector response
and background modeling need to be fine-tuned to eventually enable
line shape deconvolution in order to extract source location information
along the line of sight.
% with \about 6 Ms of exposure.
\end{abstract}
%%%%%%%%%%%%%%%%%%%%%%%%%%%%%%%%%%%%%%%%%%%%%%%%%%%%%%%%%%%%%%%%%%%%%%%%%%%

\section{Introduction}
%%%%%%%%%%%%%%%%%%%%%%%%%%%%%%%%%%%%%%%%%%%%%%%%%%%%%%%%%%%%%%%%%%%%%%%%%%%
\begin{figure}
\centering
\includegraphics[width=0.95\linewidth]{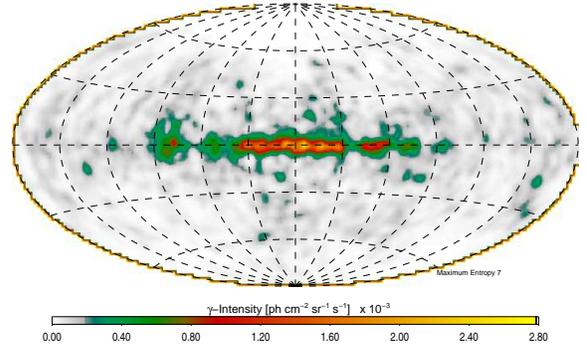}
\caption{Skymap of \Al gamma-rays, derived from the 9-year CGRO mission with 
COMPTEL \citep{plue01}}
\label{fig_almap}
\end{figure}
%%%%%%%%%%%%%%%%%%%%%%%%%%%%%%%%%%%%%%%%%%%%%%%%%%%%%%%%%%%%%%%%%%%%%%%%%%%
Radioactive \Al traces recent nucleosynthesis activity
throughout the Galaxy through gamma-rays emitted with their decay after 
a lifetime $\tau$=1.04~My. The gamma-ray photon energy is 
1808.65~($\pm$0.07)~keV 
in the laboratory frame. Such gamma-rays penetrate interstellar matter
 easily and reach gamma-ray telescopes even from within dense molecular
clouds. Their densities are $\rho\leq$10$^{-18}$g~cm$^{-3}$, the line-of-sight 
mass column
reaches $\int{\rho\hspace{3pt} ds} \ll$~g~cm$^{-2}$. Therefore interstellar clouds 
occult sources at almost all other frequencies of the electromagnetic spectrum;
only stellar envelopes reach column densities above a few g\,cm$^{-2}$, the
optical depth for such gamma-rays, so that radioactivities inside stars
and the early supernova and nova phases remain hidden from direct observations.

All-sky mapping of \Al gamma-rays with the COMPTEL imaging telescope
 (see Fig. \ref{fig_almap})
\citep{ober97,plue01} has convincingly shown that the \Al sky reflects
recent nucleosynthesis over the entire Galaxy, rather than peculiar interstellar 
\Al in the solar vicinity; the latter had been suggested from 
\Al enrichments of the early solar system as inferred from meteorites. 

%%%%%%%%%%%%%%%%%%%%%%%%%%%%%%%%%%%%%%%%%%%%%%%%%%%%%%%%%%%%%%%%%%%%%%%%%%%
%\begin{figure}
%\centering
%\includegraphics[width=0.95\linewidth]{al_grains.eps}
%\caption{Isotopic ratio measurements of meteoritic inclusions identify
%those as presolar. \Al has been inferred for presolar grains which can
%be assigned to AGB stars, core-collapse supernovae, and novae, from
%their isotopic patterns in Si, C, N, and Ti \citep{zinn98,nitt96,amar01}.}
%\label{fig_algrains}
%\end{figure}
%%%%%%%%%%%%%%%%%%%%%%%%%%%%%%%%%%%%%%%%%%%%%%%%%%%%%%%%%%%%%%%%%%%%%%%%%%%
Presolar grains in meteorites have shown in recent years
that all candidate sources of \Al (novae, AGB stars, Wolf-Rayet stars, 
and core-collapse supernovae) indeed produce \Al \citep{amar01,nitt96}.
%(see Fig. \ref{fig_algrains}). 
Yet, we consider plausible that massive 
stars dominate the Galactic \Al budget \citep{pran96}. This arises 
from the non-homogeneous 1809~keV emission along the plane of the 
Galaxy \citep{dieh95,knoe99i} reminiscent of massive-star tracers, 
and from consistency of massive-star \Al sources with dust heating 
and interstellar-medium ionization \citep{knoe99,knoe99m}. Probably
AGB stars and novae are minor contributors, unless only a subset of
them are important, which would spatially coincide with massive star regions.

%%%%%%%%%%%%%%%%%%%%%%%%%%%%%%%%%%%%%%%%%%%%%%%%%%%%%%%%%%%%%%%%%%%%%%%%%%%

\section{Status of \Al Studies}
%%%%%%%%%%%%%%%%%%%%%%%%%%%%%%%%%%%%%%%%%%%%%%%%%%%%%%%%%%%%%%%%%%%%%%%%%%%
\begin{figure}
\centering
\includegraphics[width=1.0\linewidth]{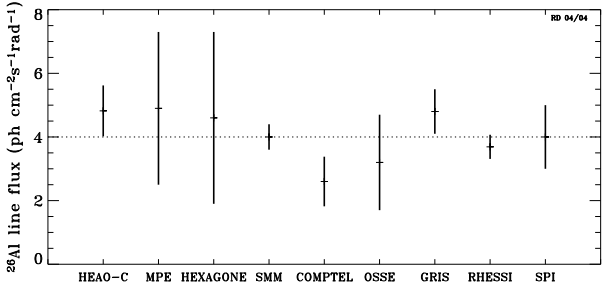}
\caption{Gamma-ray flux measurements for the inner Galaxy from different
experiments. Not all are significantly positive measurements, as seen
from their total uncertainties (1$\sigma$). 
A plausible value appears to be 4$\pm$1~10$^{-4}$ph~cm$^{-2}$s$^{-1}$ (dotted).}
\label{fig_alflux}
\end{figure}
%%%%%%%%%%%%%%%%%%%%%%%%%%%%%%%%%%%%%%%%%%%%%%%%%%%%%%%%%%%%%%%%%%%%%%%%%%%
A number of experiments have measured cosmic \Al gamma-rays.
%, mostly studying the bright ridge (Fig. \ref{fig_almap}) of the inner Galaxy. 
As the spatial resolutions of instruments vary between allsky except Earth's
shadow (RHESSI, $\approx$230$^o$),
$\approx$180$^o$ (SMM),  and $\approx$3-4$^o$ (SPI/INTEGRAL, COMPTEL), 
flux values can be compared only for large regions such as the bright 
ridge (Fig.\ref{fig_almap}) of the inner Galaxy (defined as the
inner radian, $\approx\pm$30$^o$; 
the latitude integration range is less critical, $\approx\pm$15$^o$
see Fig.\ref{fig_almap}). 
Fig.\ref{fig_alflux} summarizes the flux values which have been derived
for this region from 9 experiments, 5 of them with reported significant
detections. The discrepancies among experiments may indicate instrumental
systematics, but partly also may be due to instrument field-of-view differences
(see e.g. \citet{wund04}) 
and the definition of background references. Therefore, from this integrated
flux of $\approx$4$\pm$1~10$^{-4}$ph~cm$^{-2}$s$^{-1}$ only weak constraints
on the sources of \Al can be derived when integrated yields per type of
source from models are compared to this value. Interpretations depend on 
the spatial distribution in the Galaxy, from models smoothly following interstellar
gas as represented by exponential disks with typical scales R$\approx$3.5~kpc and 
z$\approx$0.1~kpc \citep{ferr01,dieh96}, to irregular models due to spiral
structure and peculiar, active regions such as Cygnus (see Fig. \ref{fig_almap}
and \citet{dieh96,knoe99m}).
From COMPTEL, a total Galactic mass of $\approx$2$\pm$1~M$_\odot$ has been
estimated \citep{knoe96,knoe99m}. 
All candidate sources all have been reported to be able to produce such amounts
\citep{timm95,meyn97,forr97,star00}. But yields for AGB stars are 
probably most uncertain \citep[0.01--4~\msol,][]{mowl00}, and
nova yields in current models are 0.1--0.4~\msol \citep{jose97} and would 
reach the observed total only under extreme mixing assumptions or if scaled 
with the ejected-mass ratios between models
and nova observations to obtain such high values; 
therefore these two candidate-source
arguments are somewhat controversial. For massive stars in their
Wolf-Rayet and core-collapse supernova stages, much detailed modeling is
available, and both are compatible with these inferred 2~M$_\odot$ of \Al 
in the Galaxy  (though on the low side) \citep{pran96,vuis04,raus02}.

Other arguments can be derived from a comparison of the \Al map with
tracers of the candidate sources. From comparisons of 
heated-dust distribution, free-electron distribution, and H$_\alpha$ maps,
an \Al origin in massive stars has been found most plausible
\citep{pran96, chen95, dieh96, knoe99m}. In particular, the ionizing
power and \Al yields from massive stars have been shown to be consistent: 
the Galactocentric gradient for supernova and Wolf-Rayet star
 space densities predicts an \Al yield profile as a result of different metalicity
 dependencies for these source types \citep{pran96}, 
 which matches the \Al map profile more closely
 for a Wolf-Rayet origin \citep{knoe99}; furthermore, the presence of significant
\Al in the young Cygnus massive-star region with probably-low supernova 
history suggests that the Wolf-Rayet phase of massive stars dominates
over the contribution from core collapses \citep{knoe02}.

But, detailed proof is not available yet. Convincing arguments could be
derived from better estimates of the distances and locations of candidate
sources. This can be achieved in localized source regions such as Cygnus,
Vela, and Orion (see contributions by Kn\"odlseder et al., and Schanne et al.,
these Proceedings, and \citet{dieh01}); population-synthesis modeling of the 
\Al content in the region can be compared to the gamma-ray measurements,
to check for consistency of the age of the stellar population with 
nucleosynthesis models \citep[see][]{cerv00,plue00,knoe04}. 
Moreover, high-resolution spectroscopy
together with spatial resolution on the scale of degrees, such as available 
with INTEGRAL \citep{wink03}, may allow exploitation of Galactic rotation to locate source
regions in the inner Galaxy (see below).

%%%%%%%%%%%%%%%%%%%%%%%%%%%%%%%%%%%%%%%%%%%%%%%%%%%%%%%%%%%%%%%%%%%%%%%%%%%
\begin{figure}
\centering
\includegraphics[width=1.0\linewidth]{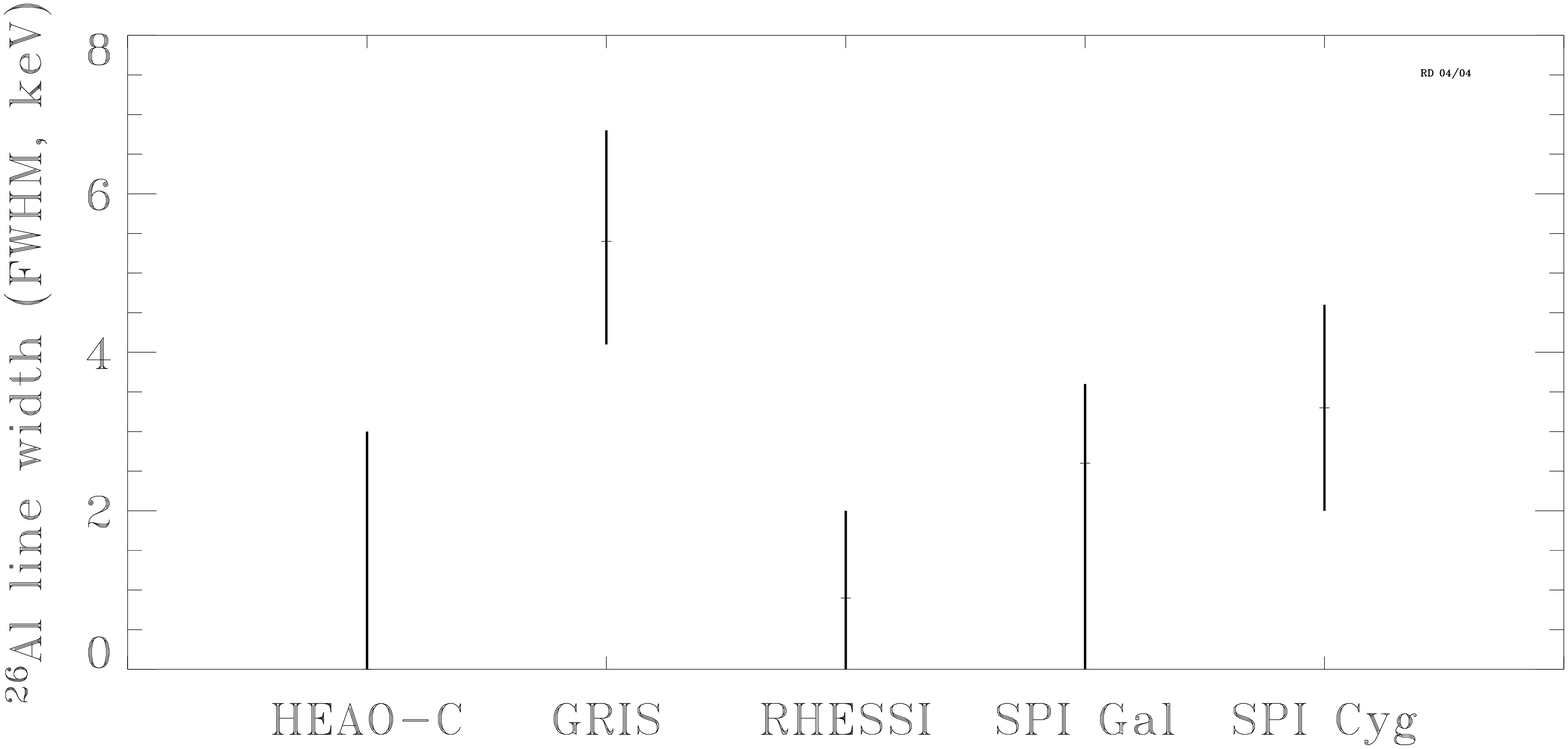}
\caption{Measurements/constraints of the intrinsic, astrophysical width of
the 1809 keV \Al line:  For the inner Galaxy three different
experiments have been made,  SPI also reported a measurement for the Cygnus region.
The significant broadening reported by GRIS is not seen by other experiments.}
\label{fig_alwidth}
\end{figure}
%%%%%%%%%%%%%%%%%%%%%%%%%%%%%%%%%%%%%%%%%%%%%%%%%%%%%%%%%%%%%%%%%%%%%%%%%%%
The measurements of the \Al gamma-ray line profile with experiments
featuring high-resolution spectroscopy (see Fig.\ref{fig_alwidth})
are somewhat controversial. The GRIS balloon-borne experiment had 
obtained a significantly-broadened line, corresponding to interstellar
velocities of decaying \Al nuclei of 540~km~s$^{-1}$ \citep{naya96};
this had been difficult to understand, unless large (kpc-sized) cavities in 
the interstellar medium or dust grains would hold the bulk of \Al 
\citep[e.g.][]{chen97,stur99}. But all satellite-based Ge detector experiments
report \Al line widths consistent with instrumental resolutions or only
slightly broadened \citep{maho82,smit03,dieh03,knoe04} 
(see Fig. \ref{fig_alwidth}); this
suggests some unknown systematic distortion of the GRIS measurement.
Still, gamma-ray spectroscopy in space at the 0.1~keV level is an
experimental challenge; careful assessments of instrumental response and
background are essential.
%%%%%%%%%%%%%%%%%%%%%%%%%%%%%%%%%%%%%%%%%%%%%%%%%%%%%%%%%%%%%%%%%%%%%%%%%%%
%%%%%%%%%%%%%%%%%%%%%%%%%%%%%%%%%%%%%%%%%%%%%%%%%%%%%%%%%%%%%%%%%%%%%%%%%%%

\section{INTEGRAL Spectrometer Data and Analysis}
The SPI spectrometer on INTEGRAL features a 19-element Ge detector camera
embedded in a massive shield of BGO detectors, with imaging capability
due to a tungsten coded-mask mounted in the telescope aperture \citep{vedr03,roqu03}.
At 1809 keV, the SPI spatial resolution is estimated as 2.8$^o$, 
point source sensitivity as 2.5~10$^{-5}$\flux\ for a narrow line, 
and spectral resolution has been measured as 3.0 keV
\citep{roqu03,jean03}. 
Spectral resolution varies over time, due to cosmic-ray irradiation of the
detectors which degrades the charge collection efficiency; this is rectified
periodically through detector annealing procedures, restoring the quoted
resolution \citep{roqu03}. Spectral analysis must account for these temporal
variations to enable line shape studies.

%%%%%%%%%%%%%%%%%%%%%%%%%%%%%%%%%%%%%%%%%%%%%%%%%%%%%%%%%%%%%%%%%%%%%%%%%%%
\begin{figure}
\centering
\includegraphics[width=0.8\linewidth]{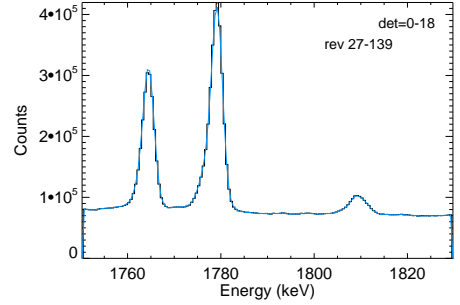}
\caption{The instrumental background spectrum in the vicinity of the \Al
line is dominated by continuum. But a feature from several blended lines
at 1810 keV, although much weaker than adjacent background lines from
Al and Bi activation at 1779 and 1764 keV, respectively, complicates
spectroscopy of the \Al line.}
\label{fig_alrawspec}
\end{figure}
%%%%%%%%%%%%%%%%%%%%%%%%%%%%%%%%%%%%%%%%%%%%%%%%%%%%%%%%%%%%%%%%%%%%%%%%%%%

%%%%%%%%%%%%%%%%%%%%%%%%%%%%%%%%%%%%%%%%%%%%%%%%%%%%%%%%%%%%%%%%%%%%%%%%%%%
\begin{figure}
\centering
\includegraphics[width=0.8\linewidth]{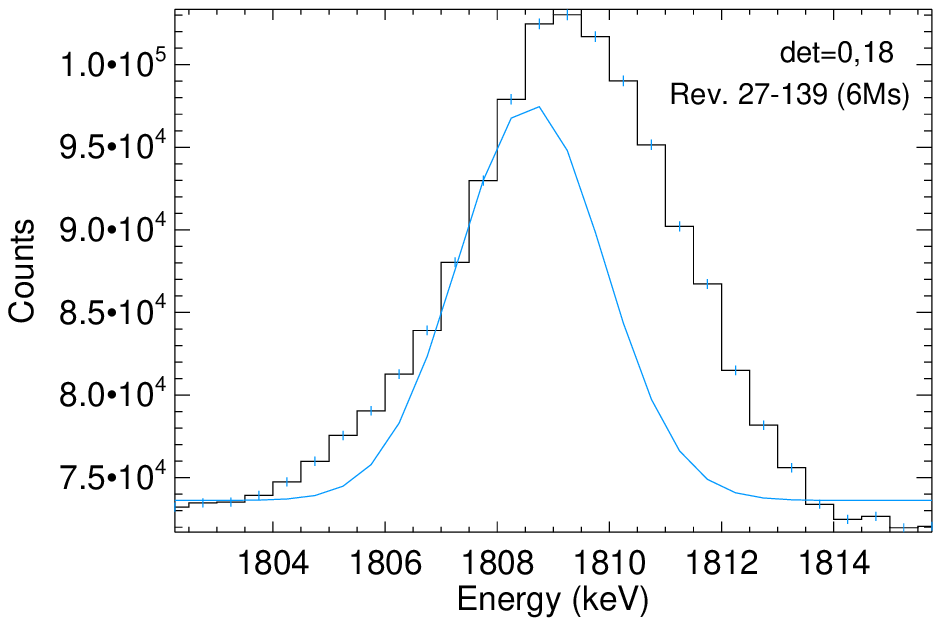}
\includegraphics[width=0.8\linewidth]{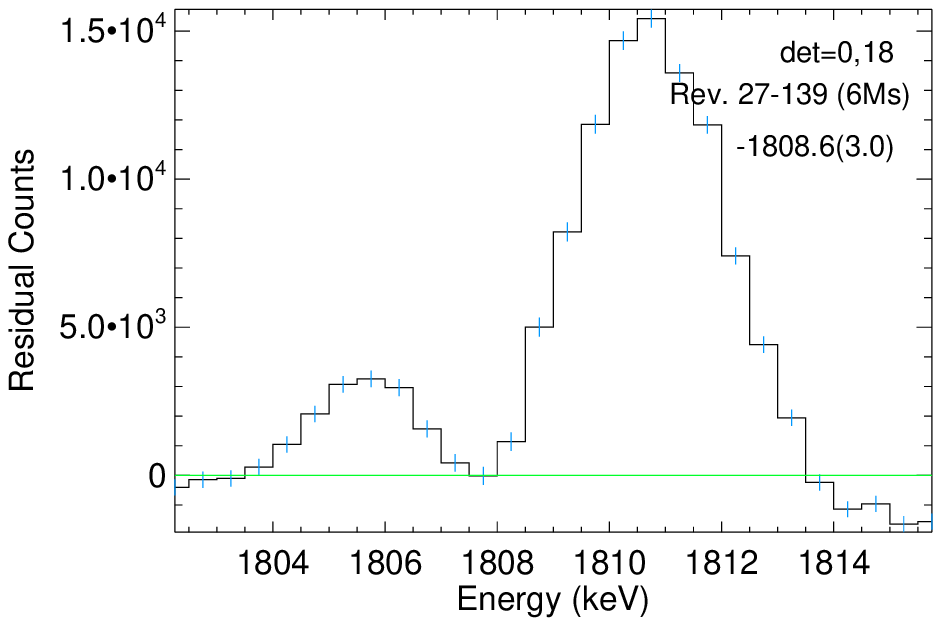}
\caption{The instrumental background feature underlying the \Al
line is probably composed of 3 lines, at energies 1805.5, 1808.7, and 1810.7 keV,
respectively. Here we show how a line at 1808.6 keV with instrumental
resolution is embedded in this feature (top). The residual spectrum (bottom)
indicates the two other instrumental line components at 1805 and 1811 keV,
both represented well by a Gaussian of instrumental width. Line intensity
ratios are 0.09/0.52/0.39, for the 1805.5, 1808.7, and 1810.7 keV lines, respectively,
with an uncertainty $\pm$0.015 in each of these.}
\label{fig_alresspec}
\end{figure}
%%%%%%%%%%%%%%%%%%%%%%%%%%%%%%%%%%%%%%%%%%%%%%%%%%%%%%%%%%%%%%%%%%%%%%%%%%%

%%%%%%%%%%%%%%%%%%%%%%%%%%%%%%%%%%%%%%%%%%%%%%%%%%%%%%%%%%%%%%%%%%%%%%%%%%%
\begin{figure}
\centering
\includegraphics[width=0.8\linewidth]{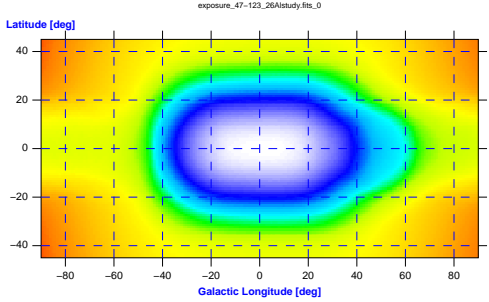}
\caption{The exposure in the inner Galaxy from GCDE parts 1 and 2, with data from
orbits 47-123. In the GC, 1.03~Ms are obtained, while in the plane
of the Galaxy, exposure is 0.52 and 0.58 Ms at -/+~30$^o$, respectively (linear scale).}
\label{fig_exposure_gcde12}
\end{figure}
%%%%%%%%%%%%%%%%%%%%%%%%%%%%%%%%%%%%%%%%%%%%%%%%%%%%%%%%%%%%%%%%%%%%%%%%%%%

The SPI sensitivity is constrained by the underlying instrumental background.
In the vicinity of the \Al line (see Fig.~\ref{fig_alrawspec}), a broad spectral feature is observed, which
probably is a blend of 3 background lines.
Background is expected at
1808.65 keV from excited $^{26}$Mg produced from spallation of Al and
from $\alpha$-captures on Na, and at 1810.7~keV from 
$^{56}$Mn($\beta^-$)$^{56}$Co($EC$)$^{56}$Fe \citep{weid03}. 
But other nuclear lines may contribute: In particular, there is a strong
hint for another, yet unidentified line, at 1805.5 keV 
(see Fig. \ref{fig_alresspec}).
The total event rate in the instrumental feature is 2~10$^{-1}$~Hz,
%slightly dominated (63\%) by single-detector events (SE); this 
which compares to an 
expected celestial \Al event rate of a few~10$^{-3}$~Hz at 1808.6 keV from 
diffuse emission in the inner Galaxy.
% (with $\approx$50\% each in SE and ME event types).  

%%%%%%%%%%%%%%%%%%%%%%%%%%%%%%%%%%%%%%%%%%%%%%%%%%%%%%%%%%%%%%%%%%%%%%%%%%%
%%%%%%%%%%%%%%%%%%%%%%%%%%%%%%%%%%%%%%%%%%%%%%%%%%%%%%%%%%%%%%%%%%%%%%%%%%%

\section{SPI Results and Issues}
\begin{figure}
\centering
\includegraphics[width=0.8\linewidth]{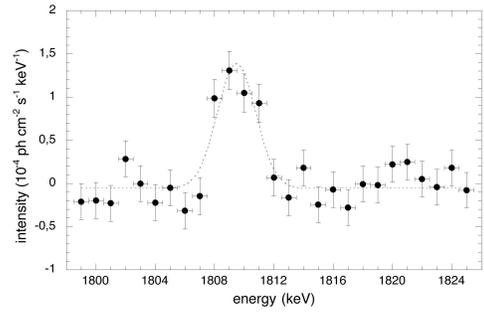}
\caption{Spectrum of fitted skymap intensities, for the earlier GCDE part-1 data
\citep[from][]{dieh03}.
 }
\label{fig_spec_obsfit_gcde1}
\end{figure}

 %%%%%%%%%%%%%%%%%%%%%%%%%%%%%%%%%%%%%%%%%%%%%%%%%%%%%%%%%%%%%%%%%%%%%%%%%%%
%\begin{figure}
%\centering
%\includegraphics[width=0.8\linewidth]{sp_res_gcde12_crab102_SPIline.eps}
%\caption{Residual count spectrum for 3.6~Ms of GCDE; the background spectrum
%used for subtraction was taken from Crab observations taken in between the
%GCDE observations, normalized to minimize residuals in the background line
%from $^{27}$Al(n,$\gamma$) at 1779 keV. A line at 1808.7~keV
%energy with instrumental resolution and 14300 counts is shown for comparison,
%and appears consistent with the data.}
%\label{fig_spec_onoff}
%\end{figure}
%%%%%%%%%%%%%%%%%%%%%%%%%%%%%%%%%%%%%%%%%%%%%%%%%%%%%%%%%%%%%%%%%%%%%%%%%%%
From the first part of the inner-Galaxy deep 
exposure (GCDE), the detection
of the 1809 keV gamma-ray line from \Al could be reported  
 \citep[][from $\approx$1~Ms of exposure selected as useful at the time]{dieh03}.
Considerable uncertainties had to be faced, since underlying background is
large and variable. Different approaches to spectral analysis and background
treatment are essential at this early stage of the mission
 to get estimates of systematic uncertainties and the
overall consistency of results, because the coded-mask
encoding of the signal is rather weak for diffuse emission. 
The \Al sky intensity from the inner $\pm$30$^\circ$ of the Galaxy
was derived as  (3--5)$ \times 10^{-4}$~ph~cm$^{-2}$s$^{-1}$,  
the line width was found consistent with SPI's instrumental
 resolution of 3~keV (FWHM). Figure \ref{fig_spec_obsfit_gcde1}
 shows a spectrum obtained from these data through one of the 
 analysis approaches discussed in \citet{dieh03}. 
 An offset by $\leq$ 0.3~keV in line energy was 
 attributed to the preliminary energy calibration.
 
In the meantime, the exposure of the inner Galaxy has been more than doubled,
and data corruption during processing have been minimized, so that 
an effective exposure (live time) of 3.6~Ms is obtained 
(see Fig. \ref{fig_exposure_gcde12}). 
Additionally, the  energy calibrations have been improved \citep{lonj04}, 
and the background behavior has been studied in much more
detail; several approaches to model the instrumental lines and continuum
are being employed and refined \citep{weid03,jean03}. 
We usefully exploit the performance record of SPI over mission time for 
energy calibration; in addition, background modeling benefits from such
continuous data, and in particular from exposures off the plane of the Galaxy 
which are presumably free of celestial \Al signal. (For the earlier analysis,
 only off-source data from Crab and LMC exposures could be used, because
of several changes of operational mode in the early part of the mission.)
With now $\approx$ 4~Ms of Galactic Plane exposure from the INTEGRAL Core
Program (Fig.~\ref{fig_exposure_gcde12}), 
and more than 2~Ms of additional performance data at other times,
the background history can be traced much more accurately.

%An intuitive way to derive a spectrum of Galactic diffuse emission
%would be to subtract a properly normalized off-source spectrum.
%This did not show any significant residuals from GCDE part 1; with more 
%data and improved energy
%calibrations, residuals hint at a signal in the right region, around
%the \Al line energy (see Fig. \ref{fig_spec_onoff}). 
%This appears re-assuring that more complex analysis
%methods are on track. 

%The next level of complexity can be introduced through normalization of
%the background, by a contraint on the relative detector signal amplitudes
%within the 19 detectors of the Ge camera. These are, in general, characteristic
%for their photon sources, e.g. isotropically-incident background photons
%will leave homogeneous count rates, while a celestial point source will
%leave a signal only in the detector elements which are not shadowed from
%the source by tungsten elements of the coded mask; for typical background,
%signatures are in between those extremes, yet characteristic of the
%background source at a particular energy and rather independent of intensity.
%With this additional constraint, subtracting off-source data with their
%characteristic amplitude pattern yields residuals shown in 
%Fig. \ref{fig_spec_onoff_detratio}. 
%
%%%%%%%%%%%%%%%%%%%%%%%%%%%%%%%%%%%%%%%%%%%%%%%%%%%%%%%%%%%%%%%%%%%%%%%%%%%
%%%%%%%%%%%%%%%%%%%%%%%%%%%%%%%%%%%%%%%%%%%%%%%%%%%%%%%%%%%%%%%%%%%%%%%%%%%
%%%%%%%%%%%%%%%%%%%%%%%%%%%%%%%%%%%%%%%%%%%%%%%%%%%%%%%%%%%%%%%%%%%%%%%%%%%
\begin{figure}
\centering
\includegraphics[width=0.8\linewidth]{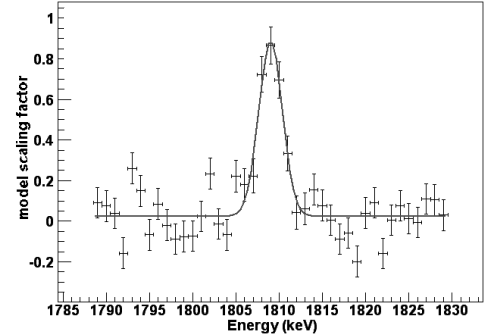}
\caption{Spectrum of fitted skymap intensities for the now-available GCDE 1+2 data.  
}
\label{fig_spec_obsfit}
\end{figure}
%%%%%%%%%%%%%%%%%%%%%%%%%%%%%%%%%%%%%%%%%%%%%%%%%%%%%%%%%%%%%%%%%%%%%%%%%%%
For an adequate model of background in the Ge detector spectra, we use
adjacent energies to force an absence of celestial signal
outside the \Al line; in the \Al line region, we adopt the spectral signature
of background from off-source measurements. As one approach, we scale this
template at fine time resolution to model line background intensity variations 
from saturated event counts in the Ge detectors as a tracer of activation,
and account for radioactive build-up where lines are identified with an isotope
\citep[see][for further details on this approach]{jean03, knoe04}.
To make use of SPI as an imaging spectrometer, we may not only impose 
constraints on characteristics expected for background, but also on the 
shadow pattern cast by
celestial photons as they illuminate the Ge camera through the coded mask; 
else the discrimination of the celestial component against
the background component will be blurred. 
We implement this through an assumption on the intensity distribution
in the sky, and fit the scaling factor of this sky model per each spectral bin,
producing as result a spectrum of celestial intensity.
Software tools {\it spidiffit} and {\it spi\_obs\_fit} are used for this 
analysis \citep[see][]{dieh03,knoe04b}.
%Here we may then adopt different constraints on the background again:
%Just assuming that the background will be at a constant level and signature
%yields a spectrum shown in Fig. \ref{spec_diffit_constbgd}. 
These efforts improve the sensitivity to the celestial signature substantially,
beyond on/off methods.
In Fig. \ref{fig_spec_obsfit} we illustrate how additional data and better 
calibrations and background treatment improve spectra for the inner Galaxy;
the analysis approach shown here is closest to the one which produced the result
shown in Fig.\ref{fig_spec_obsfit_gcde1}. 

The distribution of
\Al emission is modeled here by the emission of dust as observed by COBE/DIRBE
at 240~$\mu$m, background is modeled from adjacent energy bands adjacent to the line
(1786-1802 and 1815-1828 keV), plus a line complex template extracted from off-source
observations of the LMC (Fig. \ref{fig_spec_obsfit_gcde1}), 
and of all observations available at latitudes 
$\geq$20$^o$ (Fig. \ref{fig_spec_obsfit}), respectively, always normalized following
the intensity of saturated events in Ge detectors.
Again, variations of sky models do not affect results, but background 
models have a significant impact. Biases are possible which suppress
Galactic \Al or which still retain a contribution from
the structured instrumental background.  
Therefore, quantitative results for the Galactic large-scale \Al emission 
significantly beyond our earlier paper \citep{dieh03}, 
%such as indiacted in Fig. \ref{fig_spec_obsfit}, 
in particular intensity and gamma-ray line shape details, 
will have to await sufficiently deep understanding
of the biases of each of our analysis approaches, through
simulations and through tests on other source cases. Fine-tuning to optimize 
performances of each approach, plus uncertainty estimates per approach are 
in progress, so that consistency of these \Al results can be demonstrated.

 %%%%%%%%%%%%%%%%%%%%%%%%%%%%%%%%%%%%%%%%%%%%%%%%%%%%%%%%%%%%%%%%%%%%%%%%%%%
\begin{figure}
\centering
\includegraphics[width=0.8\linewidth]{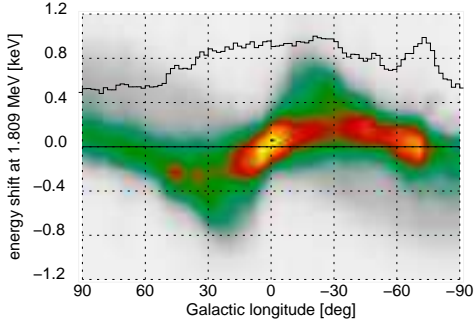}
\caption{Simulated \Al line profile along the plane of the Galaxy, from
the Doppler shifts of Galactic rotation and expansion from the ejection
events (here assumed as supernovae)\citep{kret03}. The massive-star source
distribution is shown as histogram.}
\label{fig_lineshapemap}
\end{figure}
%%%%%%%%%%%%%%%%%%%%%%%%%%%%%%%%%%%%%%%%%%%%%%%%%%%%%%%%%%%%%%%%%%%%%%%%%%%

Further sophistication of such analysis would be the generation of \Al sky intensity
distributions on the sky. This becomes feasible, once the celestial signal
is sufficiently significant so that it can be broken down into different
"pixels" on the sky. First steps towards this have been taken by splitting the
sky intensity models for different quadrants of the Galaxy, and by the
iterative imaging methods adapted from COMPTEL analyses \citep{knoe99i,stro03}.
Publication agreements within INTEGRAL's Science Working Team have led to
a split of studies of different Galactic quadrants, pursued by different subgroups.
These spatially-resolved \Al results will thus be published elsewhere. 
Most sensitive large-scale spatially-resolved studies require one to exploit all
pointings along the Galactic plane, and model background accurately around
all these observing times. 
 
\begin{figure}
\centering
\includegraphics[width=0.8\linewidth]{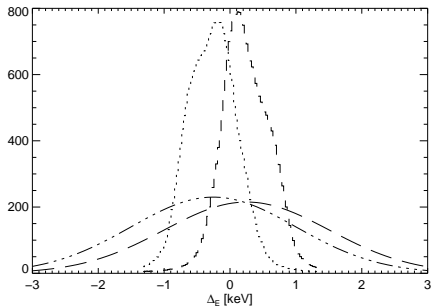}
\caption{Line profiles from regions east (dotted) and west (dashed) of the 
Galactic Center show an effective displacement of line centroids by 0.5 keV.
SPI's instrumental energy resolution blurs these idealized profiles 
(dash-dotted, long-dashed lines), so that the statistical precision of 
3~Ms of exposure is needed over these regions.}
\label{fig_lineshape_ew}
\end{figure}

Once this will be achieved, the Doppler motion of source regions at different
locations within the Galaxy is expected to leave a characteristic imprint
on the \Al gamma-ray line shape. Simulations have been performed to 
demonstrate capabilities of SPI observations \citep{kret03, kret04}.
Fig.\ref{fig_lineshapemap} shows the idealized simulated line shape (centroid, intensity,
width) versus Galactic longitude from \Al ejected by nucleosynthesis events
into a supernova-like environment \citep[see][for details]{kret03}, 
adopting current Galactic rotation models and a spatial distribution model of
sources derived from pulsar dispersion measurements, which model
free electrons in the Galaxy \citep{tayl93}. It has been shown that the ionization from 
massive stars produces a spatial distribution reminiscent of observed \Al gamma rays.
The ionization power and the \Al yields from
the population of massive stars throughout the Galaxy make up a consistent description
of massive-star origin for \Al \citep{knoe99}. 
If sufficiently large regions on the sky with Doppler motions in opposite directions
are integrated and compared against each other, a displacement of the line centroids
as large as 0.5~keV should result. With SPI's energy resolution of 3~keV, this
will be significant for a statistical accuracy corresponding to 3~Ms of exposure
throughout these regions of the Galactic plane, assuming that systematic uncertainties
are negligible in comparison (see Fig. \ref{fig_lineshape_ew}). This appears 
achievable over the duration of the INTEGRAL mission, yet presents a significant 
challenge for data analysis in the light of current uncertainties.

%%%%%%%%%%%%%%%%%%%%%%%%%%%%%%%%%%%%%%%%%%%%%%%%%%%%%%%%%%%%%%%%%%%%%%%%%%%
%%%%%%%%%%%%%%%%%%%%%%%%%%%%%%%%%%%%%%%%%%%%%%%%%%%%%%%%%%%%%%%%%%%%%%%%%%%

\section{Summary}
The Galactic emission from \Al provides unique insight into the massive star
population of the Galaxy and its interaction with the interstellar medium.
Several experiments have set the stage for addressing the related astrophysical
questions with INTEGRAL's spectrometer: the Galaxy-wide flux measurement still
lacks the precision required for source type discrimination. Local intensity
measurements such as in the Cygnus region will be a valuable complement to this
study, because in this case source populations
are better constrained than in the inner Galaxy \citep[see][]{knoe04}.
Fine spectroscopy promises to become a diagnostic, both for source distributions
within the Galaxy (through Galactic rotation), and for ejection dynamics and
the interstellar medium surrounding the sources in localized regions.
SPI measurements of the first 1.5 years have demonstrated that 
INTEGRAL has the potential to provide answers within the time frame of its mission.

{\it Acknowledgements}

%The INTEGRAL project is supported by government grants in all member states
%of the hardware teams (see http://astro.estec.esa.nl/Integral/about.html).
%The SPI project 
SPI has been completed under the responsibility and leadership of CNES.
We are grateful to ASI, CEA, CNES, DLR, ESA, INTA, NASA and OSTC for support.
We acknowledge the collaboration with David Smith, discussing
RHESSI and SPI results prior to publication. 
%\end{acknowledgements}
%%%%%%%%%%%%%%%%%%%%%%%%%%%%%%%%%%%%%%%%%%%%%%%%%%%%%%%%%%%%%%%%%%%%%%%%%%%

\end{document}